\documentstyle[aps,preprint,graphicx,amsmath,amssymb]{revtex}

\begin{document}
\title{How well can we predict CP asymmetry in $\boldsymbol{B\to \pi\pi}$, $\boldsymbol{K\pi}$ decays?}

\author{A. I. Sanda\footnote{e-mail: sanda@eken.phys.nagoya-u.ac.jp}
 and Kazumasa Ukai\footnote{e-mail: ukai@eken.phys.nagoya-u.ac.jp}}

\address{Department of Physics, Nagoya University, Nagoya, 464-8602, Japan}

\maketitle

\begin{picture}(0,0)(-355,-130)
 \put(0,0){hep-ph/0109001}
 \put(0,-20){DPNU-01-21}
\end{picture}

\begin{abstract}
 Using the perturbative QCD amplitudes for $B\to \pi\pi$ and $B\to K\pi$, 
 we have performed an extensive study of the parameter space 
 where the theoretical predictions for the branching ratios are  
 consistent with recent experimental data.
 From this allowed range of parameter space, 
 we predict the mixing induced CP asymmetry for 
 $B \to \pi^+\pi^-$ with about $11\%$ uncertainty and the other CP
 asymmetries for $B\to \pi\pi$, $K\pi$ with $40\% \sim 47\%$
 uncertainty.
 These errors are expected to be reduced as we restrict the parameter
 space by studying other decay modes and 
 by further improvements in the experimental data.
\end{abstract}

\bigskip

PACS index : 13.25.Hw, 11.10.Hi, 12.38.Bx
% 13.25.Hw Decays of bottom mesons
% 11.10.Hi Renormalization group evolution of parameters
% 12.38.Bx Perturbative calculations
% 13.25.Ft Decays of charmed mesons

%------------------------------------------------------------------------------
%                             Introduction
%------------------------------------------------------------------------------
\section{Introduction}

The mixing induced CP asymmetry for $B \to J/\psi K_S$ has been shown
to depend on only the weak phase $\phi_1$ and there is no uncertainty
from hadronic matrix elements\cite{Carter:1981tk}. 
$B$ factory is expected to yield information, not only on 
$B \to J/\psi K_S$ asymmetry but also on various other $B$ meson decays.
It has been predicted that 
$B\to K\pi$, $\pi\pi$ decay modes may have large CP
asymmetries\cite{Keum:2000ph,Lu:2000em}.
While branching ratios for these modes are very sensitive to the input
parameters, 
CP asymmetries are expected to be less sensitive.
We report here the sensitivity of predicted CP asymmetries in
the parameter region restricted by experimental values for the
branching ratios.

The time dependent CP asymmetry for $B^0(\overline{B}^0) \to f$ 
transition, where $f$ is CP eigenstate, is given by\cite{Quinn:2000jb}: 
\begin{equation}
 a_f(t) = \frac{(|\lambda_f|^2 -1) \cos (\Delta M t) + 2\, {\rm Im} \lambda_f \sin (\Delta M t)}{|\lambda_f|^2 +1},
\end{equation}
where $\lambda_f = (q/p) \overline{\rho} (f)$, and $\overline{\rho} (f)$ is
defined by the ratio of the decay amplitudes, 
$\overline{\rho} (f) = A(\overline{B} \to f)/A(B \to f)$.
For the sake of convenience, we denote the direct CP asymmetry
proportional to $\cos (\Delta M t)$ by $a_{dir}(f)$ and the mixing
induced CP asymmetry proportional to $\sin (\Delta M t)$ 
by $a_{mix}(f)$, 
\begin{equation}
 a_{dir}(f) = \frac{|\lambda_f|^2 -1}{|\lambda_f|^2 +1}, \quad 
 a_{mix}(f) = \frac{ 2\, {\rm Im} \lambda_f }{|\lambda_f|^2 +1}.
 \label{eq:def_cpasym}
\end{equation}
In general, a decay amplitude has two kinds of contributions: 
so-called tree amplitude; and penguin amplitude. 
The decay amplitude for $B \to f$ can be written as
\begin{equation}
 A(B \to f) = \xi_T T - \xi_P P,
\end{equation}
where $\xi_{T, P}$ are Kobayashi-Maskawa(KM) matrix elements, and $T, P$ are 
decay amplitudes with strong final state interaction phases for tree and
penguin contributions, respectively.
For example, $\xi_T = V_{ub}^* V_{ud}$, $\xi_P = V_{tb}^* V_{td}$ for 
$B^0 \to \pi^+\pi^-$, and $\xi_T = V_{ub}^* V_{us}$, $\xi_P = V_{tb}^*
V_{ts}$ for $B^0 \to K^+\pi^-$. 
Defining CP transformation by 
$CP |B \rangle = |\overline{B} \rangle $ and  
$CP |f \rangle = \eta_f |f \rangle $, 
the decay amplitude for the charge conjugated mode, $\overline{B} \to f$, 
can be written as
\begin{equation}
 A(\overline{B} \to f) = \eta_f (\xi_T^* T - \xi_P^* P).
\end{equation}
Thus the general expression for $\lambda_f$ is given as 
\begin{equation}
 \lambda_f = \eta_f\, e^{i 2\phi}
\cdot \frac{ 1 - r_\xi\, r_{\!\it amp}\, 
e^{- i\, {\rm arg}(\xi_P/\xi_T) }}
{ 1 - r_\xi\, r_{\!\it amp}\, e^{i\, {\rm arg}(\xi_P/\xi_T) }}, 
\label{eq:qprhobar}
\end{equation}
where $r_\xi \equiv |\xi_P/\xi_T|$, $r_{\!\it amp} \equiv P/T$, and 
$\phi \equiv {\rm arg }(V_{tb}^* V_{td}\, \xi_T^*)$.
In the above expression, $\phi$ and ${\rm arg}(\xi_P/\xi_T)$ 
have to be invariant under any rotation of quark phases, 
$q_j \to e^{i\alpha_j}q_j$.
CP asymmetry is classified into the following four types 
depending on the relationship between
tree and penguin contributions.
\begin{enumerate}
 \item[case 1.] 
       If $\xi_P$ has the same weak phase as $\xi_T$, ${\rm arg}(\xi_P/\xi_T) = 0$, 
       the hadronic matrix elements are canceled in $\lambda_f$.
       Then the direct CP asymmetry vanishes and the mixing induced CP
       asymmetry is strictly given by the weak phase $\phi$, 
       \begin{equation}
	a_{dir}(f) = 0, \quad a_{mix}(f) = \eta_f \sin (2 \phi).
	 \label{eq:cpasym1}
       \end{equation}
       In the $B \to J/\psi K_S$ decay mode, this is the case
       \cite{Carter:1981tk}.
       The experimental data $a_{mix}(J/\psi K_S)$ makes it possible
       to determine the weak phase 
       $\phi_1 = \pi - {\rm arg }(V_{tb}^* V_{td} V_{cb} V_{cd}^*)$
       \cite{Aubert:2001nu,Abe:2001xe}.
 \item[case 2.] 
       If the tree contribution is much larger than that of the penguin,  
       $|\xi_T T| \gg |\xi_P P|$, 
       $\lambda_f$ is expressed only by KM matrix elements just like in 
       the case 1., $\lambda_f = \eta_f e^{i 2 \phi}$.
       This leads to the same CP asymmetries as shown in 
       eq.\eqref{eq:cpasym1}.
       When there is no interference between tree and penguin
       contribution, the direct CP asymmetry does vanish, 
       and the mixing induced CP asymmetry is directly related to the 
       weak phase $\phi$.
 \item[case 3.] 
       If the penguin contribution is much larger than that of the tree, 
       $|\xi_P P| \gg |\xi_T T|$, 
       $\lambda_f$ is expressed only by the angle between $V_{tb}^* V_{td}$ 
       and $\xi_P$, 
       \begin{equation}
        \lambda_f = \eta_f e^{i\, 2{\rm arg}(V_{tb}^* V_{td}\, \xi_P^*)}.
       \end{equation}
       In this case, the direct CP asymmetry vanishes, 
       and the mixing induced CP asymmetry is directly related to some
       weak phase likewise, 
       \begin{equation}
         a_{dir}(f) = 0, \quad a_{mix}(f) = \eta_f \sin (2 \phi'), 
         \label{eq:cpasym2}
       \end{equation}
       where $\phi' \equiv {\rm arg }(V_{tb}^* V_{td}\, \xi_P^*)$.
       In the $B \to \phi K_S$, $K_S \pi^0$ decay modes, this is the case.

       (1) $B \to \phi K_S$ decay mode:\\
       There is no tree contribution in this decay mode, $T =0$.
       The weak phase $\phi'$ for this decay mode is defined by 
       $\omega_{tc}^{ds}$ in the $ds$ triangle\cite{Aleksan:1994if}, 
       \begin{equation}
       \phi' = {\rm arg}(V_{td} V_{ts}^* V_{cd}^* V_{cs})
             \equiv \omega_{tc}^{ds}.
       \end{equation}
       The CP asymmetry is strictly given by the weak phase 
       $\omega_{tc}^{ds}$, 
       \begin{equation}
        a_{mix}(\phi K_S) = -\sin(2\omega_{tc}^{ds}). 
        \label{eq:asym3}
       \end{equation}
       This makes it possible to extract the weak phase $\omega_{tc}^{ds}$ 
       from the measurement of the mixing induced CP asymmetry 
       $a_{mix}(\phi K_S)$.
       Here $\omega_{tc}^{ds}$ is related to $\phi_1$ as follows,  
       \begin{equation}
        \omega_{tc}^{ds} = - \phi_1 + \omega_{ct}^{sb}, 
        \label{eq:phisd_phi1}
       \end{equation}
       where $\omega_{ct}^{sb} \equiv {\rm arg}(V_{cs} V_{cb}^*/V_{ts} V_{tb}^*) = \pi + {\cal O}(\lambda^2)$.
       $a_{mix}(\phi K_S) \neq \sin 2(\phi_1 - \omega_{ct}^{sb})$ 
       implies the presence of new physics\cite{London:1989ph,London:1997zk}.

       (2) $B \to K_S \pi^0$ decay mode:\\
       While we realize that it is difficult to measure the time 
       dependence of $a_{K_S\pi^0}$ at current $B$ factories, 
       we will show that $B \to K_S \pi^0$ decay mode satisfies 
       $|\xi_T T|/|\xi_P P| \sim 2\times 10^{-2}$.
       Here 
       $\xi_T = V_{cd} V_{cs}^* V_{ub}^* V_{us}$,   
       $\xi_P = V_{cd} V_{cs}^* V_{tb}^* V_{ts}$, where 
       $V_{cd} V_{cs}^*$ factor comes from
       the $K^0$-$\overline{K}^0$ mixing.
       Then, the mixing induced CP asymmetry for $B \to K_S \pi^0$ decay mode
       is given by the weak phase $\omega_{tc}^{ds}$, 
       \begin{equation}
         a_{mix}(K_S \pi^0) = -\sin(2\omega_{tc}^{ds}).
         \label{eq:asym_Kspi}
       \end{equation}
       Eq.\eqref{eq:asym_Kspi} has been obtained by 
       Ref.\cite{London:1997zk}.
       But they have assumed that $B^0 \to \pi^+\pi^-$ and 
       $B^0 \to K^+\pi^-$ decay
       modes are dominated by the tree and penguin contributions, 
       respectively.
       These assumptions are inconsistent with experiment.
       Belle Collaboration has found the ratio of the
       branching ratios\cite{Abe:2001nq}, 
       \begin{equation}
        \frac{{\rm BR}(B^0 \to K^+\pi^-)}{2\, {\rm BR}(B^0 \to K^0\pi^0)}
        = 0.60^{+0.25}_{-0.29} {}^{+0.11}_{-0.16}.
        \label{eq:kpi_ratio}
       \end{equation}
       Neglecting the tree contribution, isospin analysis leads to
       a conclusion that the above ratio is equal to $1$. 
       The eq.\eqref{eq:kpi_ratio} implies that the tree contribution 
       can not be neglected in the $B^0 \to K^+\pi^-$ decay mode.
 \item[case 4.] 
       If the penguin contribution is comparable with that of the tree, 
       the direct CP asymmetry does not necessarily vanish, and the
       mixing induced CP asymmetry has impurities from the penguin or
       tree contribution. 
       $B \to \pi\pi$ and $B^0 \to K^+\pi^-$ fall into this case.

       In $B \to \pi\pi$ decay mode, 
       both KM factors, $\xi_T$ and $\xi_P$, have the same order of
       magnitude, and the tree 
       contribution can interfere with the penguin amplitude. 
       The CP asymmetry is parameterized as 
       \begin{gather}
        a_{dir}(\pi\pi) = \frac{2 r \sin \delta \sin \phi_2}
         {1 + r^2 + 2 r \cos \delta \cos \phi_2}, \\
         a_{mix}(\pi\pi) = (1- a_{dir}(\pi\pi)) {\rm Im}\!
        \left[ e^{i(2\phi_2)}
        \frac{1 + r e^{i(\delta - \phi_2)}}
        {1 + r e^{i(\delta + \phi_2)}} \right], 
        \label{eq:mix_asym}
       \end{gather}
       where $r_\xi = |V_{tb}^* V_{td}/V_{ub}^* V_{ud}|$, 
       $r \equiv r_\xi |r_{\!\it amp}|$, 
       and $\delta \equiv {\rm arg}(r_{\!\it amp})$ is the relative 
       strong phase. 
       The weak phase $\phi_2$ is defined as the angle between 
       $V_{ub}^* V_{ud}$ and $V_{tb}^* V_{td}$, 
       \begin{equation}
        \phi_2 = {\rm arg}\left( 
        \frac{V_{tb}^* V_{td}}{-V_{ub}^* V_{ud}}\right).
       \end{equation}
       Unless we know both the magnitude and phase of $r_{\!\it amp}$, 
       it is impossible to extract the weak
       phase $\phi_2$ from the data $a_{mix}(\pi\pi)$. 
       In principle, isospin analysis makes it possible to overcome such
       a pollution without understanding the penguin 
       contribution\cite{Gronau:1990ka}.
       In order to perform the isospin analysis, all modes
       for $B \to \pi\pi$ have to be measured. 
       However, it is difficult to measure the branching ratio of 
       $B^0(\overline{B}^0) \to \pi^0\pi^0$, which has background problem
       as well as a tiny branching ratio of 
       ${\cal O}(10^{-7})$\cite{Lu:2000em}.
       Therefore, in practice, it is hard to perform the isospin
       analysis.

       In the $B \to K^+\pi^-$ decay mode, 
       $\xi_T = V_{ub}^* V_{us}$ is smaller than $\xi_P = V_{tb}^*V_{ts}$ 
       by ${\cal O}(\lambda^2)$, and 
       the tree contribution does interfere with penguin. 
       For this mode, $a_{mix}(K^+\pi^-)=0$ and the direct CP asymmetry
       is given by,  
       \begin{eqnarray}
         a_{dir}(K^+\pi^-) 
         &=& \frac{\Gamma(\overline{B}^0 \to K^-\pi^+) 
                      - \Gamma(B^0 \to K^+\pi^-)}
                  {\Gamma(\overline{B}^0 \to K^-\pi^+) 
                      + \Gamma(B^0 \to K^+\pi^-)} \nonumber \\
         &=& \frac{- 2 r \sin \delta \sin \omega_{tu}^{sb}}
         {1 + r^2 - 2 r \cos \delta \cos \omega_{tu}^{sb}}, 
       \end{eqnarray}
       where $r_\xi \equiv |V_{tb}^* V_{ts}/V_{ub}^* V_{us}|$.
       The weak phase $\omega_{tu}^{sb}$ in the $sb$ triangle is defined as
       $\omega_{tu}^{sb} = {\rm arg}(V_{ts} V_{tb}^*/V_{us} V_{ub}^*)$
       \cite{Aleksan:1994if}.
       Note that 
       the $\omega_{tu}^{sb}$ is related to the weak
       phase $\phi_3$ as follows:
       \begin{equation}
         \omega_{tu}^{sb} = \pi - \phi_3 + \omega_{ct}^{sb} -\omega_{uc}^{ds},
       \end{equation}
       where $\phi_3 \equiv {\rm arg }(-V_{ub}^* V_{ud}/ V_{cb}^* V_{cd})$, 
       $\omega_{uc}^{ds} \equiv {\rm arg}(V_{ud} V_{us}^* V_{cd}^* V_{cs}) = \pi + {\cal O}(\lambda^4)$.
\end{enumerate}
We stress that, unless we know the ratio between tree and penguin
contribution with the relative strong phase, we can predict neither the
$a_{dir}$ nor $a_{mix}$.

Perturbative QCD(PQCD) approach has been developed to theoretically
understand semi-leptonic and two-body hadronic $B$ meson
decays\cite{Wu:1996gb,Keum:2000ph,Lu:2000em,Kurimoto:2001zj,pqcdworks}.
This approach enables us to calculate both the phase and magnitude 
of tree and penguin amplitudes.
In this paper, applying PQCD approach to $B \to \pi\pi$ and 
$B \to K\pi$ decay modes\footnote{In Ref.
\cite{Keum:2000ph,Lu:2000em}, we used the wave functions for
light meson which were obtained phenomenologically. 
In this paper, we consider all twist-$3$ contributions, and use the wave
functions which were decided from light-cone QCD sum rule. The updated
results hardly change the previous ones\cite{ukai1}.}, 
we predict the ratios $r_{\!\it amp}$ between the tree and penguin amplitudes, and
give CP asymmetries without relying on the isospin
analysis.

%------------------------------------------------------------------------------
%                           Numerical results
%------------------------------------------------------------------------------
\section{Numerical results}
\begin{figure}[htbp]
\begin{center}
   \includegraphics[height=7.0cm]{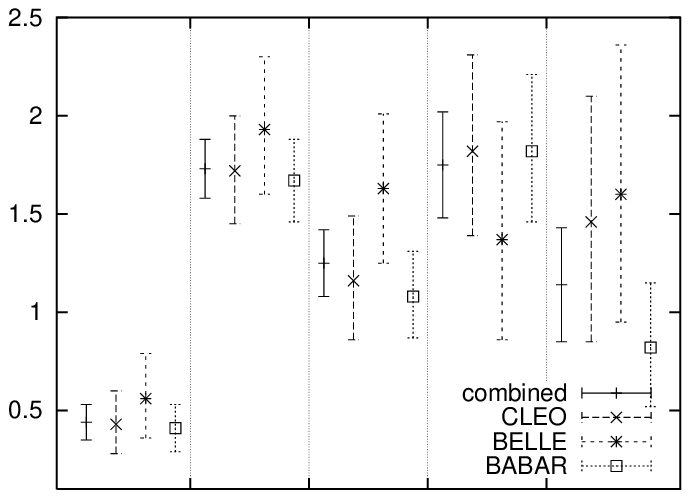}\\
  \begin{picture}(0,0)(0,0)
   \put(-100,5){$\pi^+\pi^-$}
   \put(-55,5){$K^+\pi^-$}
   \put(-10,5){$K^+\pi^0$}
   \put(35,5){$K^0\pi^+$}
   \put(80,5){$K^0\pi^0$}
   \put(-155,90){\rotatebox{90}{{\rm BR}$(10^{-6})$}}
  \end{picture}
\end{center}
\caption{The branching ratios for $B\to\pi^+\pi^-$ and $B\to K\pi$ 
measured at CLEO, Belle, and BABAR.
The combined data are also shown here.}
\label{fig:brdata}
\end{figure}

Applying PQCD approach, we take all twist-$3$ contributions into
account, and use
the wave functions for light mesons, which were decided from light-cone
QCD sum rule, 
\begin{gather}
\Phi_\pi(x) = \frac{1}{\sqrt{2N_c}} \gamma_5
\{ \not \! P_\pi \phi_\pi^A(x) 
+ m_{0\pi} \phi_\pi^P(x) 
+ m_{0\pi} (\not v_\pi \not n_\pi -1) \phi_\pi^T(x)
\}, \\
\Phi_K(x) = \frac{1}{\sqrt{2N_c}} \gamma_5
\{ \not \! P_K \phi_K^A(x) 
+ m_{0K} \phi_K^P(x) 
+ m_{0K} (\not v_K \not n_K -1) \phi_K^T(x)
\}, 
\end{gather}
where $N_c = 3$ is  color's degree of freedom, $v_{\pi(K)}^\mu$,
$n_{\pi(K)}^\mu$ are normalized to dimensionless unit vectors, and 
$v_{\pi(K)} \propto P_{\pi(K)}$, $n_{\pi(K)} \perp v_{\pi(K)}$,
$v_{\pi(K)} \cdot n_{\pi(K)} = 1$. 
$x$ is momentum fraction of light quark's momentum in the meson to
parent meson's one.
$m_{0\pi(K)}$ are defined by the quark condensate, 
\begin{equation}
m_{0\pi} = \frac{M_\pi^2}{m_u + m_d} =
-\frac{2 \langle 0 |\overline{q}q|0 \rangle}{f_\pi^2}, \quad 
m_{0K} = \frac{M_K^2}{m_u + m_s} 
= -\frac{\langle 0 |\overline{q}q + \overline{s}s |0 \rangle}{f_K^2},
\end{equation}
where they are given as\cite{Ball:1998je}, 
\begin{equation}
 m_{0\pi} = 1.6 \pm 0.2 \mbox{ GeV},\quad
 m_{0K} = 1.6 \pm 0.2 \mbox{ GeV}, 
\end{equation}
without ${\rm SU}(3)$ flavour violation.
Lorentz scalar wave functions for light mesons, $\phi_{\pi(K)}^{A,P,T}$,
are expanded by Gegenbauer polynomials,
\begin{gather}
\phi_\pi^A(x) 
= \frac{f_\pi}{2\sqrt{2 N_c}} 6 x(1-x)
\Bigl\{ 1 - a_2^\pi \cdot \frac{3}{2} (1 - 5 \xi^2)
+ a_4^\pi \cdot \frac{15}{8} (1 - 14\xi^2 +21\xi^4)
\Bigr\}, \\
\phi_K^A(x) 
= \frac{f_K}{2\sqrt{2 N_c}} 6 x(1-x)
\Bigl\{ 1 - a_1^K \cdot 3 \xi 
- a_2^K \cdot \frac{3}{2} (1 - 5 \xi^2) \Bigr\}, \\
\phi_{K(\pi)}^P(x) 
= \frac{f_{K(\pi)}}{2\sqrt{2 N_c}} 
\left\{ 1 - a_{p1}^{K(\pi)} \cdot \frac{1}{2} (1 - 3\xi^2)
+ a_{p2}^{K(\pi)} \cdot \frac{1}{8} (3 - 30\xi^2 +35\xi^4) \right\}, \\
\phi_{K(\pi)}^T(x) 
= \frac{f_{K(\pi)}}{2\sqrt{2 N_c}} (1-2x)
\left\{ 1 + a_T^{K(\pi)} \cdot 3(-3 + 5 \xi^2) \right\},
\end{gather}
where $\xi \equiv 2x -1$.
The coefficients $a^{K(\pi)}_{p1,p2,T}$ are given as function of
$m_{0K(\pi)}$, $a_2^{K(\pi)}$ and some input parameters, 
$\eta_3$, $\omega_3$ in Ref.\cite{Ball:1998je}, 
where $a_1^K$, $a_2^{K(\pi)}$, $a_4^\pi$, $\eta_3$, and $\omega_3$ 
are calculated from QCD sum rule within $30$\% accuracy.
$B$ meson's wave function is parameterized by two Lorentz scalar wave
functions, 
\begin{equation}
\Phi_B(x,b) = \frac{1}{\sqrt{2N_c}}
(\not \! P_B + M_B ) \gamma_5 [\phi_B(x,b) + 
(\not \! n_+ - \not \! n_-) \bar{\phi}_B(x,b)], 
\end{equation}
where $n_+ = (1,0,0_T)$, $n_- = (0,1,0_T)$, and 
$b$ is the relative separation between $b$-quark and light quark in the
$B$ meson.
According to Ref.\cite{Kurimoto:2001zj}, 
the contribution from $\bar{\phi}_B$ is found to be negligible, and 
we adopt as the wave function $\phi_B$ at rest, 
\begin{equation}
\phi_B(x,b) = N_B x^2(1-x)^2 \exp \left[ 
-\frac{M_B^2\ x^2}{2 \omega_B^2} -\frac{1}{2} (\omega_B b)^2
\right], 
\end{equation}
where the normalization constant $N_B$ is fixed by the decay constant
$f_B$, $\omega_B$ parameterizes the extent of $B$ meson, and 
$\omega_B$ is order of the mass difference between $B$ meson and
$b$-quark, 
$\omega_B \sim {\cal O}(\overline{\Lambda})$, 
$\overline{\Lambda} \equiv M_B - m_b$.
Note that the decay rate is very sensitive to $\omega_B$.
Now, the branching 
ratios\cite{Abe:2001nq,Cronin-Hennessy:2000hw,Aubert:2001hs}
will be used to restrict the parameter space.
The combined branching ratios are 
\begin{eqnarray}
 {\rm BR}(B^0 \to \pi^+\pi^-) &=& (0.44 \pm 0.09)\times 10^{-5}, \nonumber\\
 {\rm BR}(B^0 \to K^+\pi^-) &=& (1.73 \pm 0.15)\times 10^{-5}, \nonumber\\
 {\rm BR}(B^+ \to K^+\pi^0) &=& (1.25 \pm 0.17)\times 10^{-5}, \nonumber\\
 {\rm BR}(B^+ \to K^0\pi^+) &=& (1.75 \pm 0.27)\times 10^{-5}, \nonumber\\
 {\rm BR}(B^0 \to K^0\pi^0) &=& (1.14 \pm 0.29)\times 10^{-5}, 
\label{eq:brdata}
\end{eqnarray}
which are shown in Figure \ref{fig:brdata}.
The errors of the measured branching ratios are still
large, and it is too early to put them together. 
Therefore, 
we obtain the range of allowed parameters so that the calculated
branching ratios are consistent with the data, eq.\eqref{eq:brdata}, 
within $2\sigma$, 
where we scan the parameters in the wave function within 
$0.35 \leq \omega_B \leq 0.55$ GeV, 
$m_{0\pi} = 1.6 \pm 0.2$ GeV, $m_{0K} = 1.6 \pm 0.2$, 
and $a_1^K$, $a_2^{K(\pi)}$, $a_4^\pi$, $\eta_3$, $\omega_3$ 
given in Ref.\cite{Ball:1998je} within $30\%$ ranges.

\begin{figure}[htbp]
\begin{center}
 \vspace*{5.0cm}
 \includegraphics[height=14.0cm]{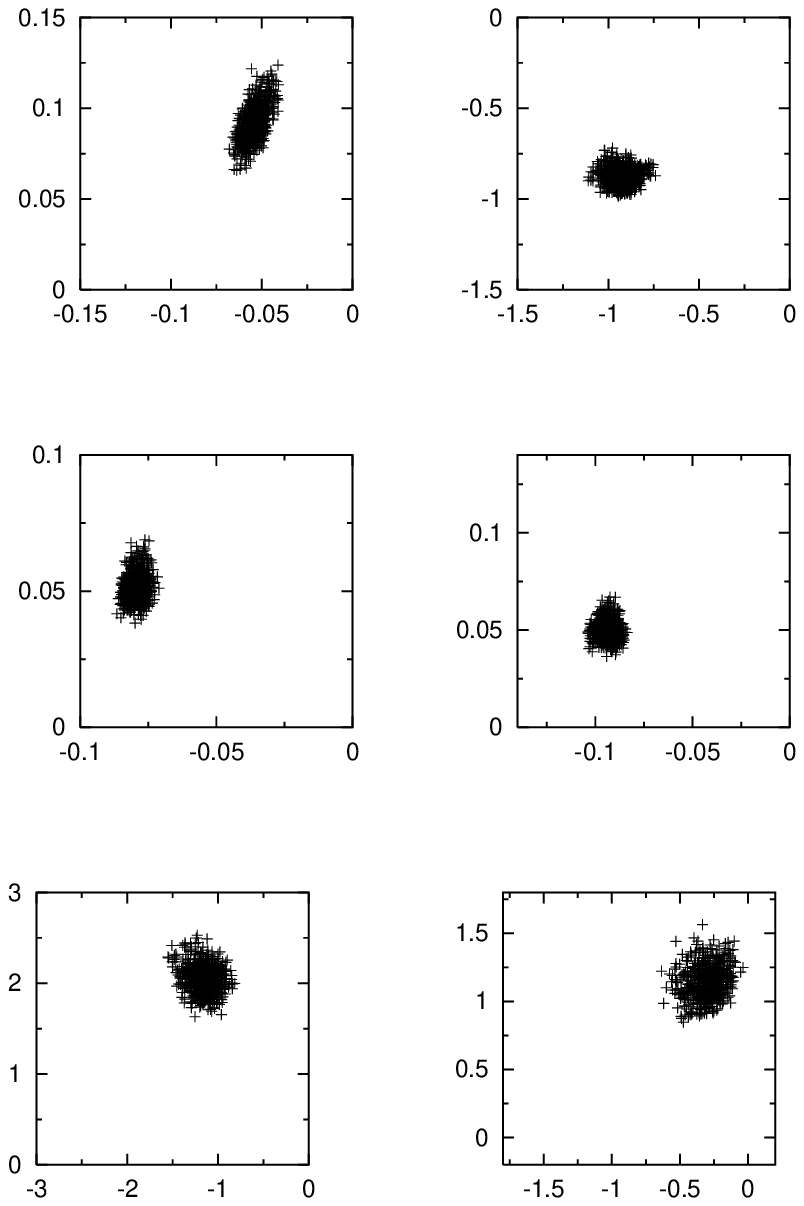}
  \begin{picture}(0,0)(40,0)
   \put(-120,405){${\rm Re}[r_{\!\it amp}(\pi^+\pi^-)]$}
   \put(-200,470){\rotatebox{90}{${\rm Im}[r_{\!\it amp}(\pi^+\pi^-)]$}}
   \put(80,405){${\rm Re}[r_{\!\it amp}(\pi^0\pi^0)]$}
   \put(5,470){\rotatebox{90}{${\rm Im}[r_{\!\it amp}(\pi^0\pi^0)]$}}
   \put(-120,205){${\rm Re}[r_{\!\it amp}(K^+\pi^-)]$}
   \put(-200,270){\rotatebox{90}{${\rm Im}[r_{\!\it amp}(K^+\pi^-)]$}}
   \put(80,205){${\rm Re}[r_{\!\it amp}(K^+\pi^0)]$}
   \put(0,270){\rotatebox{90}{${\rm Im}[r_{\!\it amp}(K^+\pi^0)]$}}
   \put(-140,5){${\rm Re}[r_{\!\it amp}(K^0\pi^+)]$}
   \put(-205,70){\rotatebox{90}{${\rm Im}[r_{\!\it amp}(K^0\pi^+)]$}}
   \put(75,5){${\rm Re}[r_{\!\it amp}(K^0\pi^0)]$}
   \put(0,70){\rotatebox{90}{${\rm Im}[r_{\!\it amp}(K^0\pi^0)]$}}
  \end{picture}
\end{center} 
\caption{The ratios $r_{\!\it amp}$ between the penguin and tree amplitude for
 $B\to \pi\pi$, $K\pi$ decay modes. We do not show here
 $r_{\!\it amp}$ for $B^+\to \pi^+\pi^0$ because of $|r_{\!\it amp}(\pi^+\pi^0)|\sim 0$.
For $r_{\!\it amp}(\pi^+\pi^-)$, the value obtained here corresponds to
$d\, e^{i\theta} = (0.24 \sim 0.36)\, e^{i(101^\circ \sim 130^\circ)}$ 
in the notation of Ref.\protect\cite{Fleischer:2000un}.
This should be compared with the value, 
$d\, e^{i\theta} = 0.09[0.18]\, e^{i\, 193[187]^\circ}$,  
given in Ref.\protect\cite{Beneke:1999br}.}
\label{fig:r}
\end{figure}

The calculated $r_{\!\it amp}$ for each final state are shown 
at Figs.\ref{fig:r}, and summarized in Table \ref{tb:rA}, 
where we do not take higher order corrections into account.
They show the error for $|r_{\!\it amp}|$, denoted as $\Delta |r_{\!\it amp}|/|r_{\!\it amp}|$, 
is $10 \sim 26\%$.
In principle, a calculation to ${\cal O}(\alpha_s^2)$ 
should be performed. 
Higher twist contributions should be studied.
These will be done in the future.
For now, we assume that these corrections will not exceed $30$\% 
in the amplitude.
The error $\Delta |r_{\!\it amp}|/|r_{\!\it amp}|$ 
is $30 \sim 40\%$ with such higher order corrections.
Because the factorizable tree amplitude has color suppression 
in $B^0 \to \pi^0\pi^0$, $K^0\pi^0$ decay modes, and the tree
amplitude comes from only annihilation diagram in 
$B^+ \to K^0\pi^+$ decay mode, 
their decay modes have $r_{\!\it amp} \sim {\cal O}(1)$.
Considering KM factor, $r_\xi \sim 2.4$ for $B\to\pi\pi$ and 
$r_\xi \sim 50$ for $B\to K\pi$, 
Table \ref{tb:rA} shows that the penguin contribution is dominant in 
$B \to K^0\pi^+$, $K^0\pi^0$ decay modes and 
comparable with that of tree in $B \to \pi^0\pi^0$, $K^+\pi^-$, and
$K^+\pi^0$ decay modes.
In PQCD, the decay amplitudes are given by integrating the convolution
of wave functions, hard amplitude, and Wilson's coefficient, over
loop momenta.
Therefore, the dependence on the wave functions does not cancel completely in
the ratio $r_{\!\it amp}$.
This leads to blobs shown in Figs.\ref{fig:r}.
Uncertainty in KM factors, decay constants, and parameters in
the wave functions exerts an influence on the branching ratios.
We elect the central values for the KM 
factors\cite{Groom:2000in,Ciuchini:2000de} and $B$ meson's decay 
constant $f_B = 200 \pm 30$ MeV\cite{saoki} in Figs.\ref{fig:r}.
Figs.\ref{fig:r} show the value of $r_{\!\it amp}$ for each set
of parameters which predict the branching ratios to within $2\sigma$ 
of eq.\eqref{eq:brdata}.

\begin{figure}[htbp]
 \begin{center}
  \includegraphics[height=4.5cm]{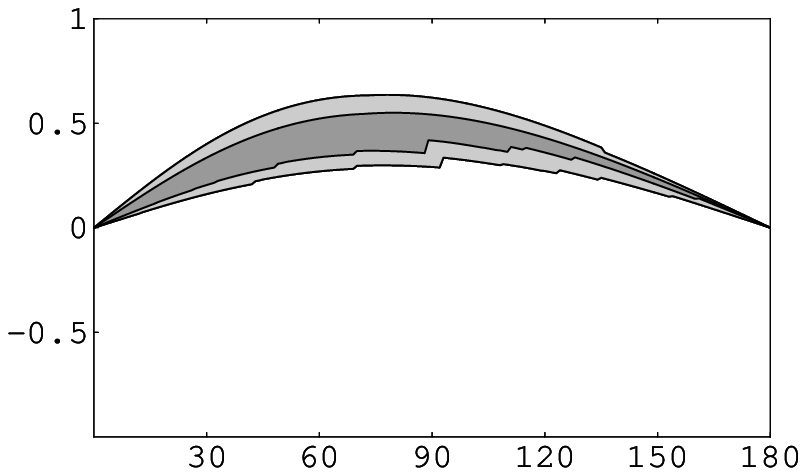}
  \vspace{1.0cm}\\
  \includegraphics[height=4.5cm]{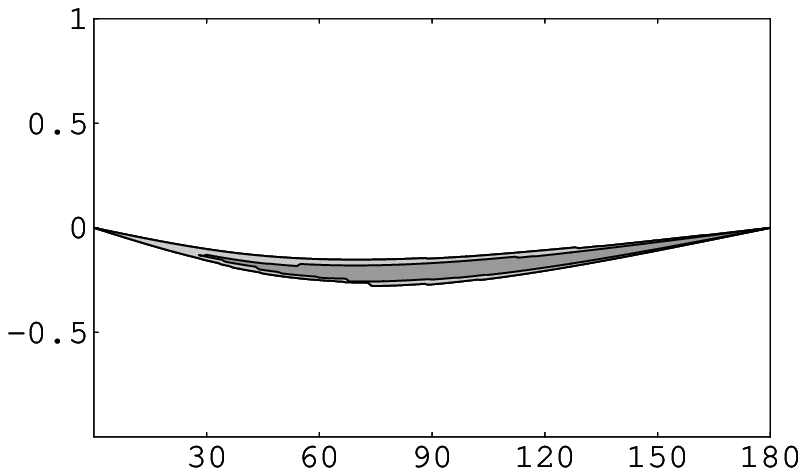}
  \hspace{0.5cm}
  \includegraphics[height=4.5cm]{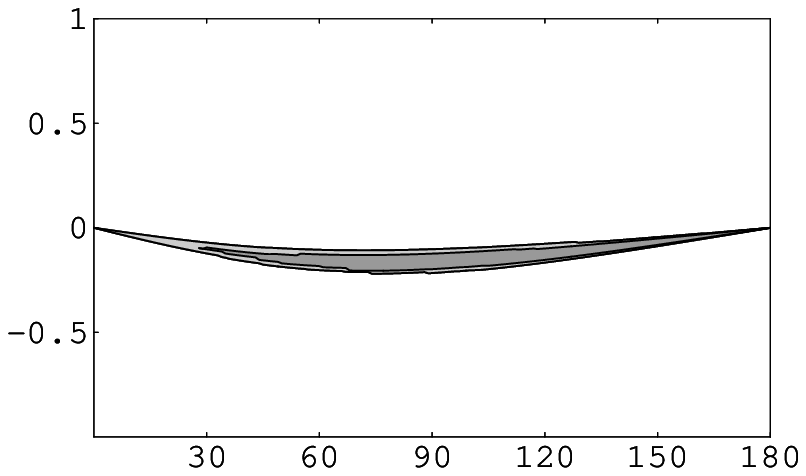}\\
  \begin{picture}(0,0)(0,0)
   \put(-15,165){$\phi_2$ (degree)}
   \put(-120,205){\rotatebox{90}{$a_{dir}(B\to \pi^+ \pi^-)$}}
   \put(-130,5){$\phi_3$ (degree)}
   \put(-235,45){\rotatebox{90}{$a_{dir}(B\to K^+ \pi^-)$}}
   \put(100,5){$\phi_3$ (degree)}
   \put(-5,45){\rotatebox{90}{$a_{dir}(B\to K^+ \pi^0)$}}
  \end{picture}
 \end{center}
 \caption{Direct CP asymmetry for $B\to \pi\pi$, $K\pi$ decay modes.
The central value of KM
 factors\protect\cite{Groom:2000in,Ciuchini:2000de}
gives the darker shaded regions, and the lighter shaded regions include
 the error of KM factors.
$a_{dir}(\pi^+ \pi^0)$ is almost zero for any $\phi_2$.
$a_{dir}(K^0 \pi^+)$ is almost zero for any $\phi_3$.
$a_{dir}(K^0 \pi^0)$ becomes maximum at $\phi_3 = 90^\circ$,
 $a_{dir}(K^0 \pi^0) = -0.04$.}
 \label{fig:dircp}
\end{figure}

\begin{figure}[htbp]
 \begin{center}
  \includegraphics[height=4.5cm]{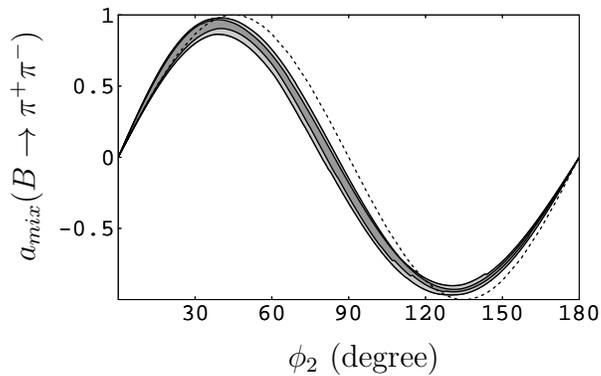}\\
  \begin{picture}(0,0)(-115,0)
   \put(-130,5){$\phi_2$ (degree)}
   \put(-235,45){\rotatebox{90}{$a_{mix}(B\to \pi^+ \pi^-)$}}
  \end{picture}
 \end{center}
 \caption{Mixing induced CP asymmetry for $B\to \pi^+\pi^-$.
The difference from the dotted line($\sin\, 2\phi_2$) shows sizeable
 penguin pollution.}
 \label{fig:cpasym_pipi}
\end{figure}

We prefer not to mix the error from the parameter space and the error
from the higher order corrections.
So, for now, we leave aside the error from the higher order corrections.
From the allowed range of $r_{\!\it amp}$, we compute the CP asymmetry using 
eqs.\eqref{eq:def_cpasym},\eqref{eq:qprhobar}.
The results are shown in Figs.\ref{fig:dircp}, \ref{fig:cpasym_pipi}.
They are summarized in Table \ref{tb:asym_dir}.
Now the higher order corrections are added to the result shown 
in Table \ref{tb:asym_dir} in quadrature.
It results in $40\sim 47$\% uncertainty for $a_{dir}$.
The measured CP asymmetries have been presented in 
Ref.\cite{Abe:2001hs,Aubert:2001qj}, where Belle Collaboration gives 
$-0.25 < {\cal A}(K^\mp \pi^\pm)\, (= a_{dir}(K^+\pi^-)) < 0.37$ at 
$90$\% confidence level, and BABAR Collaboration gives 
${\cal A}_{K\pi}\, (= a_{dir}(K^+\pi^-)) = -0.07\pm 0.08\pm 0.02$.
They are not inconsistent with our predictions.

Although the relative error of $|r_{\!\it amp}|$ 
is $10 \sim 26\%$ without higher order corrections, 
the mixing induced CP asymmetry for $B\to \pi^+\pi^-$, 
$a_{mix}(\pi^+\pi^-)$, has small uncertainty, $6\%$.
This is because the magnitude of $r_{\!\it amp}$ is small, $|r_{\!\it amp}| \sim 0.1$.
Since $a_{mix}$ has zeroth order term in expansion of $|r_{\!\it amp}|$, 
the error $\Delta a_{mix}/a_{mix}$ is proportional to $r$:
\begin{equation}
\frac{\Delta a_{mix}}{a_{mix}}(\pi^+\pi^-) = - r \sin \phi_2
(\sin\delta - \cos\delta - \cot 2\phi_2 \sin\delta)
\frac{\Delta |r_{\!\it amp}|}{|r_{\!\it amp}|} + {\cal O}(|r_{\!\it amp}|^2).
\end{equation}
The factor $r = r_\xi|r_{\!\it amp}|\sim 0.3$ leads to small uncertainty for $a_{mix}$.
The error from the higher order corrections is also suppressed by the
factor $r$, and the total error for $a_{mix}$ is estimated at $11$\%.
In contradistinction to $a_{mix}$,
the direct CP asymmetry has large uncertainty. 
Since $\Delta a_{dir}/a_{dir}$ is not suppressed by $r$, 
\begin{equation}
 \frac{\Delta a_{dir}}{a_{dir}}(\pi^+\pi^-) 
= \frac{\Delta |r_{\!\it amp}|}{|r_{\!\it amp}|} + {\cal O}(|r_{\!\it amp}|),
\end{equation}
the direct CP asymmetry is more sensitive to 
$r_{\!\it amp}$ compared to the mixing induced CP asymmetry.
Similar errors are assigned to $B\to K^+\pi^-$ and $K^+\pi^0$.

The mixing induced CP asymmetry for $B\to K_S\pi^0$, 
$a_{mix}(K_S\pi^0)$, has little uncertainty.
Here the factorizable tree amplitude is subjected to color suppression, 
and the penguin amplitude without KM factor is comparable to that of the
tree, $r_{\!\it amp} \sim 1$.
However, KM factor, 
$r_\xi = |V_{tb}^* V_{ts}/V_{ub}^* V_{us}| \sim 50$, is very large and 
it makes the penguin contribution dominant.
Since $r_\xi |r_{\!\it amp}| \gg 1$, the first term in the numerator and denominator 
can be neglected in eq.\eqref{eq:qprhobar}, and 
$\Delta |r_{\!\it amp}|$ is canceled out as shown in section 1.
This is the reason why $a_{mix}(K_S\pi^0)$ has little uncertainty.
$r_\xi |r_{\!\it amp}| \sim 50$ is classified into the case 3 in
section 1, and $a_{mix}(K_S\pi^0)$ is given by, 
\begin{equation}
 a_{mix}(K_S\pi^0) = \sin 2(\phi_1  - \omega_{ct}^{sb}), 
\end{equation}
where its uncertainty from $r_{\!\it amp}$ is $0.03\%$.

%------------------------------------------------------------------------------
%                               Conclusion
%------------------------------------------------------------------------------
\section{Conclusion}

In order to extract the weak phase from the experimental data on CP
asymmetries, we have to know the magnitude and strong phase of the
penguin contribution.
We have applied PQCD approach to $B\to \pi\pi$ and $B\to K\pi$ decay
modes, which allows us to compute both the strong phase and 
magnitude of these amplitudes.
We have calculated the ratios $r_{\!\it amp}$ between the penguin and
tree amplitude which are crucial in predicting CP asymmetries.
For each decay mode, uncertainty of $|r_{\!\it amp}|$ is $30 \sim 40\%$.
The direct CP asymmetry which is proportional to $r_{\!\it amp}$ 
suffers from uncertainty in $|r_{\!\it amp}|$ directly, 
and the error is estimated at $40\% \sim 47\%$.
However, the mixing induced CP asymmetry for $B\to \pi^+\pi^-$, 
which is not proportional to $r_{\!\it amp}$, 
has small uncertainty, and the asymmetry has been predicted with $11\%$
uncertainty.
While 
it is difficult to measure the time dependence of $a_{K_S\pi^0}$ 
at current $B$ factories, 
we predicted $r_{\!\it amp}$ for $B\to K^0\pi^0$ decay mode.
This leads to the mixing induced CP asymmetry which is related to the
weak phase $\phi_1  - \omega_{ct}^{sb}$.

There are other attempts to study CP violation in $B\to\pi\pi$, $K\pi$ 
decays\cite{Fleischer:2000un}.
In the analysis of Ref.\cite{Fleischer:2000un}, 
${\rm SU}(3)_F$ symmetry is parameterized and 
$r_{\!\it amp}$ is obtained from the analysis of
Ref.\cite{Beneke:1999br}.
Our value for $r_{\!\it amp}$ which corresponds to 
the penguin parameter 
$d\, e^{i\theta} = (0.24 \sim 0.36)\, e^{i(101^\circ \sim 130^\circ)}$ 
in the notation of Ref.\cite{Fleischer:2000un} 
seems to yield a better agreement with experiments.

Improvement in experimental data is expected in the near future.
The uncertainty in $r_{\!\it amp}$ will be also expected to improve.
The predicted CP asymmetries will make it possible to extract the weak
phase from their experimental data.
%---------------------------------------------------------------------------
%                            Acknowledgments
%---------------------------------------------------------------------------
\section*{Acknowledgments}
We would like to thank PQCD group members: 
H.-W. Huang, Y.Y. Keum, E. Kou, T. Kurimoto, H.-n. Li, C.-D. L\"u,
S. Mishima, N. Sinha, R. Sinha, M.-Z. Yang and T. Yoshikawa
for fruitful discussions.
We are grateful to Y. Okada and M. Hazumi for discussions. 
Research of K.U. is supported by the Japan Society for 
 Promotion of Science under the Predoctoral Research Program. 
This work is supported 
by Grand-in Aid for Special Project Research (Physics of CP violation), 
by the Grant-in-Aid for Science Research, Ministry of Education, Science
and Culture, Japan(No.12004276).
%------------------------------------------------------------------------------
%                               references
%------------------------------------------------------------------------------

%------------------------------------------------------------------------------
%                                figures
%------------------------------------------------------------------------------
%------------------------------------------------------------------------------
%                                tables
%------------------------------------------------------------------------------
\begin{table}[htbp]
 \begin{center}
  \begin{tabular}[tb]{ccc}
   final state & $|r_{\!\it amp}|$ & ${\rm arg}(r_{\!\it amp})$\\
   \hline
   $\pi^+\pi^-$ & $0.089 \sim 0.134$ & $108^\circ \sim 135^\circ$ \\
   $\pi^0\pi^0$ & $1.11 \sim 1.43$ & $-144^\circ \sim -130^\circ$ \\
   $\pi^+\pi^0$ & $0.0113 \sim 0.0146$ & $180^\circ$ \\
   $K^+\pi^-$ & $0.087 \sim 0.106$ & $137^\circ \sim 155^\circ$ \\
   $K^+\pi^0$ & $0.095 \sim 0.117$ & $143^\circ \sim 159^\circ$ \\
   $K^0\pi^+$ & $1.91 \sim 2.85$ & $112^\circ \sim 128^\circ$ \\
   $K^0\pi^0$ & $0.95 \sim 1.60$ & $92^\circ \sim 122^\circ$
  \end{tabular}
 \end{center}
 \caption{$r_{\!\it amp}$ for each final state calculated in PQCD. 
Uncertainty comes from parameters in the wave functions.}
 \label{tb:rA}
\end{table}
\begin{table}[htbp]
 \begin{center}
  \begin{tabular}[tb]{cccc}
    final state & $\phi_i$ & $a_{dir}$ & $\Delta a_{dir}/a_{dir}$ \\
   \hline
   $\pi^+\pi^-$ & $80^\circ$ & $0.366 \sim 0.550\ (0.298 \sim 0.636)$ 
   & $20\%\ (36\%)$ \\
   $\pi^+\pi^0$ & any & $\sim 0$ & \\
   $K^+\pi^-$ & $70^\circ$ & $-0.257 \sim -0.180\ (-0.261\sim -0.153)$ 
   & $18\%\ (26\%)$ \\
   $K^+\pi^0$ & $70^\circ$ & $-0.205 \sim -0.130\ (-0.211\sim -0.108)$ 
   & $22\%\ (32\%)$ \\
   $K^0\pi^+$ & any & $\sim 0$ & \\
   $K^0\pi^0$ & $90^\circ$ & $-0.0393\sim -0.0237\ (-0.0428\sim -0.0213)$ 
   & $25\%\ (33\%)$ \\
   \hline
   \hline
    & $\phi_2$ & $a_{mix}$ & $\Delta a_{mix}/a_{mix}$ \\
   \hline
   $\pi^+\pi^-$ & $40^\circ$ & $0.904 \sim 0.966\ (0.864\sim 0.979)$ 
   & $3.3\%\ (6.2\%)$ \\
  \end{tabular}
 \end{center}
 \caption{
For illustration purposes, we show direct CP asymmetries and mixing
 induced CP asymmetry for $B\to \pi\pi$, $K\pi$ decay modes.
We have chosen $\phi_i$ ($\phi_i= \phi_2$ for $B\to \pi\pi$ and 
$\phi_i = \phi_3$ for $B \to K\pi$) which give the maximum asymmetries.
The parenthesized values include errors of KM factors,
while the unparenthesized values are obtained from the central values of KM
 factors\protect\cite{Groom:2000in,Ciuchini:2000de}.
The higher order corrections are not considered in the above errors.
}
 \label{tb:asym_dir}
\end{table}

\end{document}